\documentclass[aps,twocolumn,nofootinbib,
preprintnumbers,showkeys,
superscriptaddress]{revtex4-2}

\usepackage{graphicx}
\usepackage{dcolumn}
\usepackage{bm}
\usepackage{amsmath}
\usepackage{wasysym}
\usepackage{float}
\usepackage{multirow}
\usepackage{xcolor}
\usepackage{scrextend}
\usepackage[colorlinks=true, pdfstartview=FitV, bookmarks=true, bookmarksnumbered=true, breaklinks]{hyperref}
\usepackage{color}
\definecolor{blue}{rgb}{0.0, 0.0, 1.0}
\definecolor{red}{rgb}{1.0, 0.0, 0.0}
\definecolor{royalblue}{rgb}{0.0, 0.14, 0.4}
\hypersetup{linkcolor=royalblue, citecolor=blue, urlcolor=royalblue}
\newcommand{\mytitle}[1]{\vspace{.5cm}{\em #1.---}}
\def\orcid#1{\kern .08em\href{https://orcid.org/#1}{\includegraphics[keepaspectratio,width=0.7em]{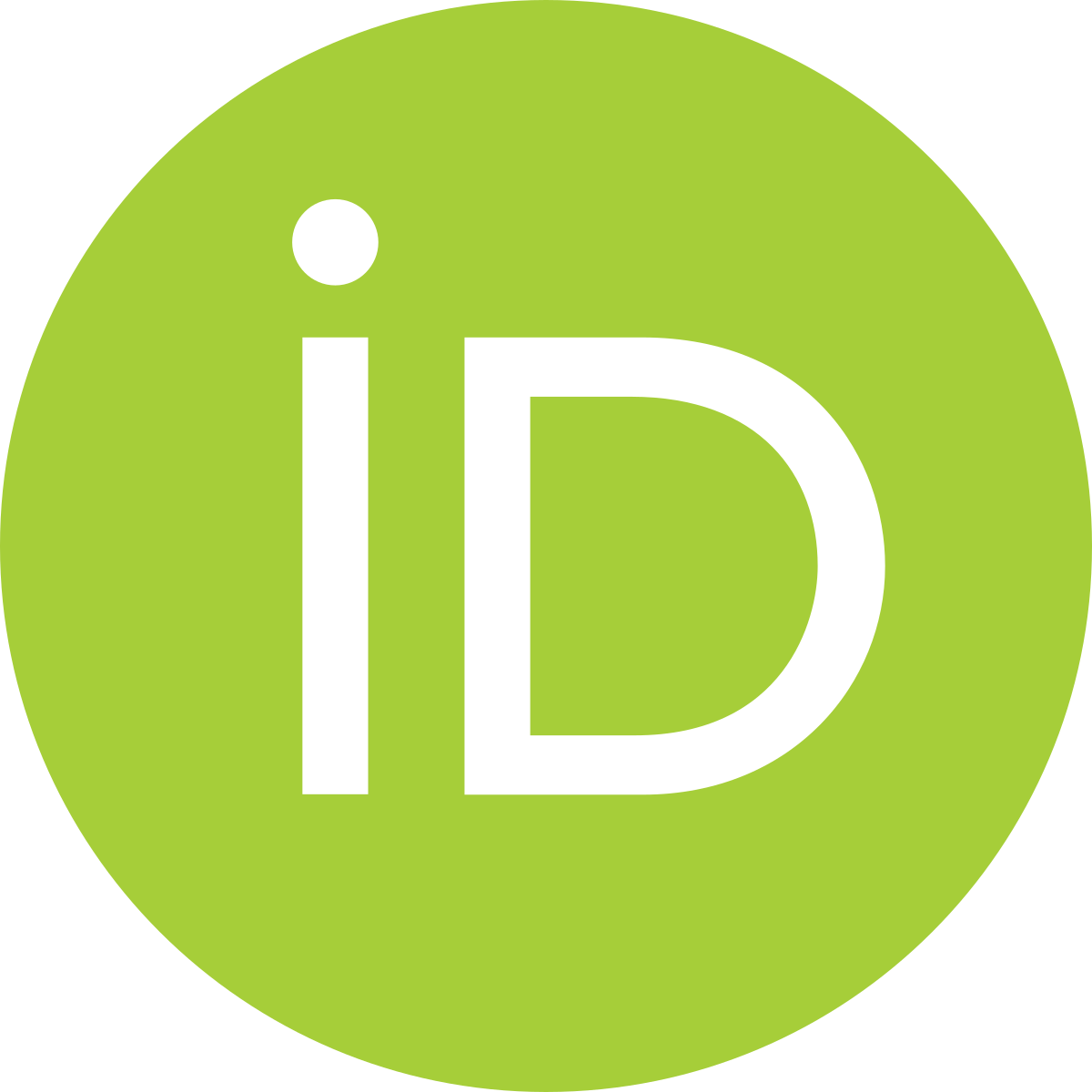}}}

\begin{document}
\title{Unveiling the pole structure of S-matrix using deep learning}

\author{Denny Lane B. Sombillo\orcid{0000-0001-9357-7236}}
\email[]{sombillo@rcnp.osaka-u.ac.jp}
\email[]{dbsombillo@up.edu.ph}
\affiliation{National Institute of Physics, University of the Philippines Diliman, Quezon City 1101, Philippines}
\affiliation{Research Center for Nuclear Physics (RCNP), Osaka University, Ibaraki, Osaka 567-0047, Japan}
%\orcid{0000-0001-9357-7236}
\author{Yoichi Ikeda\orcid{0000-0002-2235-1464}}
\affiliation{Department of Physics, Kyushu University, Fukuoka 819-0395, Japan}
%\orcid{0000-0002-2235-1464}
\author{Toru Sato\orcid{0000-0001-5216-5657}}
\affiliation{Research Center for Nuclear Physics (RCNP), Osaka University, Ibaraki, Osaka 567-0047, Japan}
%\orcid{0000-0001-5216-5657}
\author{Atsushi Hosaka\orcid{0000-0003-3623-6667}}
\affiliation{Research Center for Nuclear Physics (RCNP), Osaka University, Ibaraki, Osaka 567-0047, Japan}
\affiliation{Advanced Science Research Center, Japan Atomic Energy Agency, Tokai, Ibaraki 319-1195, Japan}
%\orcid{0000-0003-3623-6667}

\date{\today}

\begin{abstract}
	Particle scattering is a powerful tool to unveil the nature of various subatomic phenomena.  The key quantity is the scattering amplitude whose analytic structure carries the information of the quantum states.  
	In this work, we demonstrate our first step attempt to extract the pole configuration of inelastic scatterings using the deep learning method.
	Among various problems, motivated by the recent new hadron phenomena, we develop a curriculum learning method of deep neural network to analyze coupled channel scattering problems.  
	We show how effectively the method works to extract the pole configuration associated with resonances in the $\pi N$ scatterings.    
\end{abstract}		
\maketitle

\mytitle{Introduction}The past decades saw the proliferation of new peak structures in hadronic scattering cross-sections with properties eluding the conventional quark model~\cite{Olsen2018,Pc2019,dijpsi2020,DongBaruGuoHanhartNefediev2020,Wang:2019evy,Haidenbauer:2021smk}. 
Specifically, most of these peaks appear close to some two-hadron threshold, leading to a variety of possible interpretations of its nature. 
Some enhancements are claimed to be candidates of the conventional $qqq$ or $q\bar{q}$ resonant excitation~\cite{QuarkModels2000,Jaffe2007}, or one of the novels multi-quark states like pentaquarks or tetraquarks~\cite{Ali2017,Multiquark2019}.  
The proximity of peaks to the two-hadron threshold suggests that the coupled channel interaction in the $S$-wave of two hadrons plays a significant role~\cite{Threshold2021}. 
For instance, the observed peaks may be generated by the two-hadron interaction, which may lead to either a hadronic molecule or a virtual state ~\cite{Guo2018,Yamaguchu2020,Hyodo2013}. 
On the other hand, peaks can also be interpreted as kinematical threshold effects where the presence of a dynamical physical state is not relevant. 
These kinematical effects can be attributed to either the two-body threshold cusp or the three-body triangle singularity~\cite{TriangleOset2016,TriangleAndCusp,SXNakamura2020}. 
Identifying which of the observed near-threshold peaks correspond to resonances is crucial to our ongoing quest to understand the mechanism of color confinement.

Whether due to resonance or enhanced two-body threshold cusp, peak structures are generally associated with the presence of nearby S-matrix poles. In a coupled-channel scattering, these poles can be on any of the multiple unphysical energy sheets. Knowing which sheet is occupied by the poles can give us useful information on the nature of observed peaks. Further analyses can proceed by either using a model-dependent or model-independent approach.

For the model-dependent treatment, the pole position is obtained by fitting the parameters of a model to the experimental data. 
In a coupled-channel analysis, the model contains some channel coupling parameters that can be switched off. This model's feature allows us to determine the nature of enhancement by tracing the origin of the pole in the zero-coupling limit~\cite{Frazer1964,PearceGibson,Badalyan1982,Hanhart2014}. 
Alternatively, a model-independent analysis is possible by using the effective range expansion around some threshold of interest. 
Specifically, a pole-counting method was proposed in Refs.~\cite{MorganPennington1991,Morgan1992} which relates the number of near-threshold poles on different unphysical sheets to the nature of enhancement. 
It was later shown in~\cite{Baru2004} that the pole-counting method is consistent with Weinberg's compositeness criterion~\cite{Weinberg1965}. 
It follows that the identification of pole configuration, which we define to be the number of poles in each sheet associated with the enhancement, is a crucial starting point in a model-independent analysis of near-threshold phenomena. 
The task is now reduced to a classification problem.
For this purpose, the deep learning approach is one of the best ways to be employed. Here, we show how to extract the pole configuration of the S-matrix using deep learning.

Deep learning is a versatile tool used in the physical sciences~\cite{MLandPS}. The availability of a large dataset is essential to improve the performance of a deep neural network (DNN) model. 
Unlike in other fields, experimental data in hadron physics is limited and with statistical uncertainties.
It is, therefore, imperative to introduce a general method to generate the teaching dataset containing a large number of simulated amplitudes with known pole configurations. Here, the simulated amplitudes are mock data generated from the general properties of the S-matrix. The generation of teaching dataset is followed by the construction and improvement of the DNN model, which is then applied to the experimental data.
We have demonstrated in Ref.~\cite{Sombillo2020} the feasibility of using deep learning in the analysis of single-channel scattering where a generic S-matrix is used to generate the teaching dataset. Here, we extend our method to the coupled-channel case.

\begin{figure}[ht!]
	\centering
	\includegraphics[width=\columnwidth]{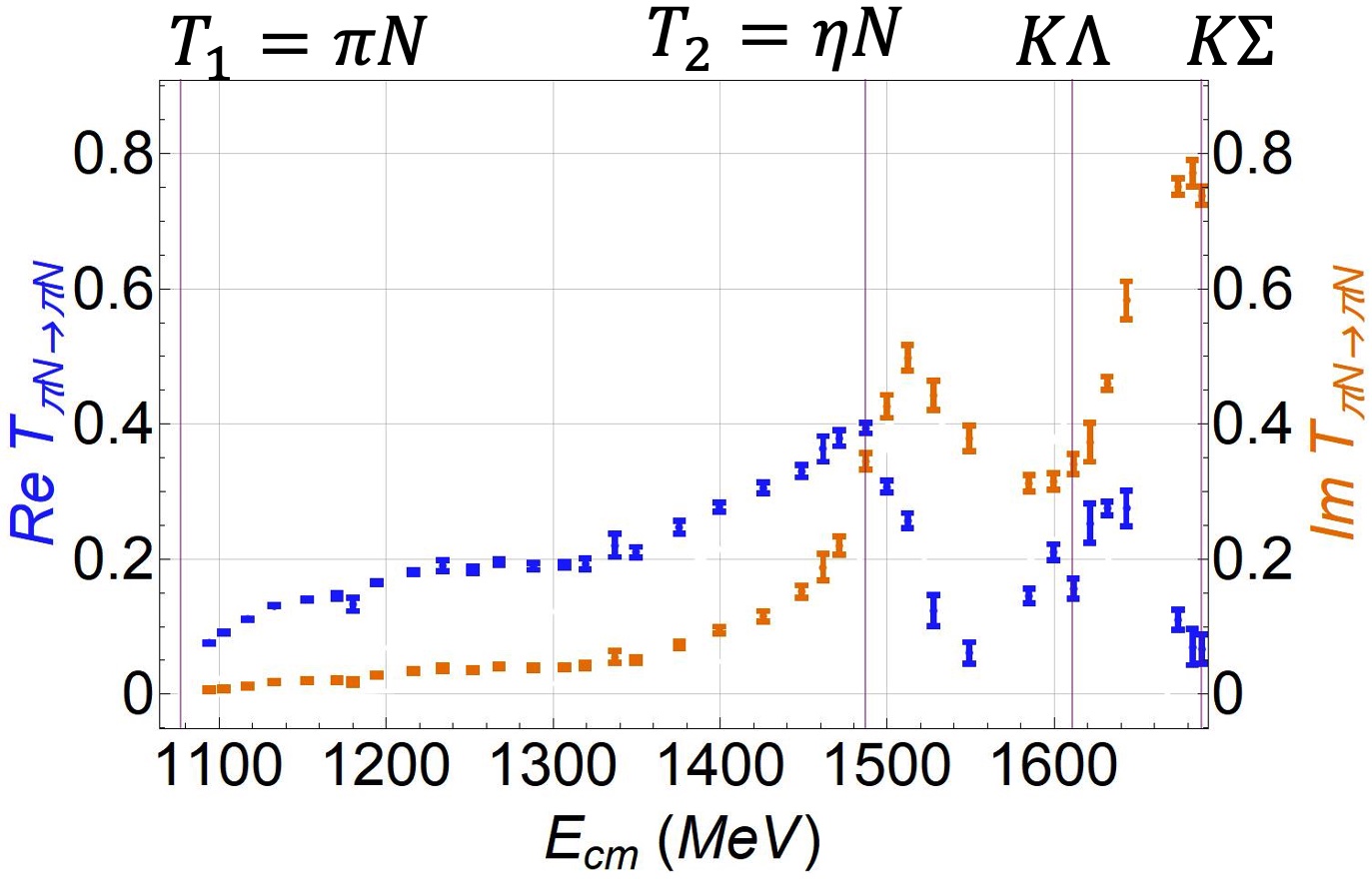}
	\caption{The elastic $\pi N$ amplitude of the GW-SAID in Ref.~\cite{GW_SAID}. 
		The two-hadron thresholds are shown as vertical thin lines. 
		Only the $\pi N$ and $\eta N$ channels are considered in the present study. 
		The $K\Lambda$ and $K\Sigma$ thresholds are shown for reference purposes only.}
	\label{fig:SAID_piN}
\end{figure}

\begin{figure*}[ht!]
	\includegraphics[width=\linewidth]{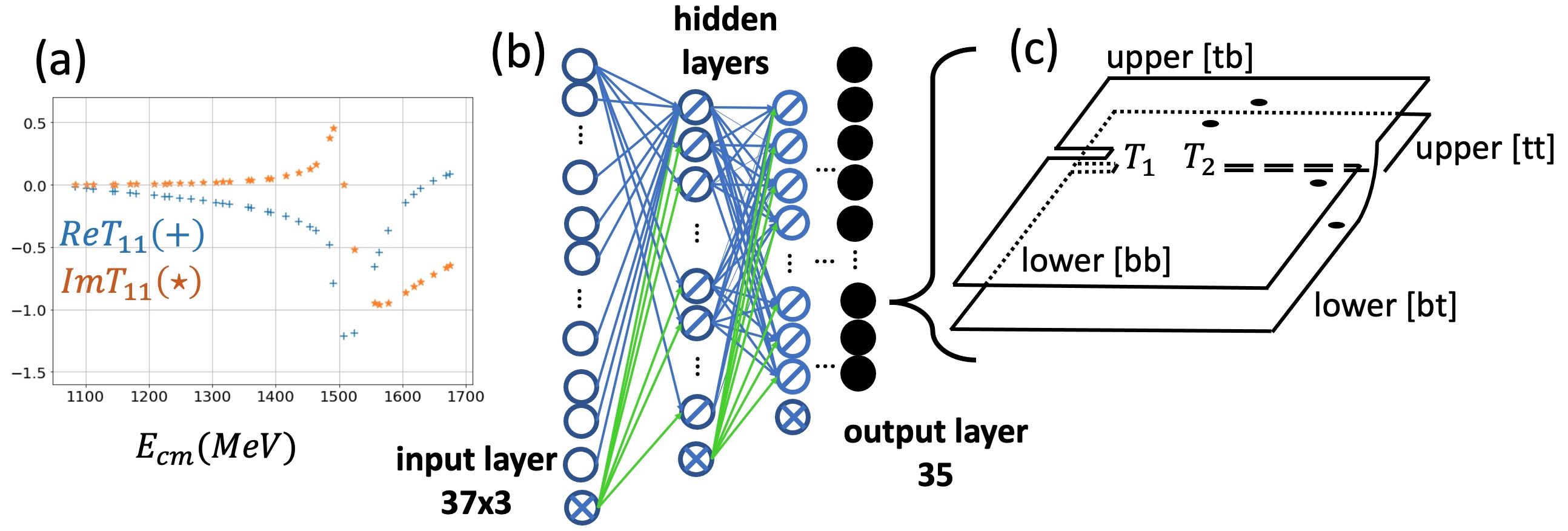}
	\caption{Schematic operation of our DNN model where (a) is one of the simulated amplitudes (b) is the DNN model and (c) is one of the representative output pole configuration. For (b), $\Circle$s are for input nodes, $\otimes$s are for bias nodes $\oslash$s are for hidden layer nodes and $\CIRCLE$s are for output nodes. The dashed lines in (c) are connected to each other.}
	\label{fig:architecture}
\end{figure*}

In this letter, we present a deep learning approach to study the coupled-channel scattering with near-threshold enhancement. 
We design a classification DNN that determines the S-matrix pole configuration when an input partial-wave amplitude is given. 
In doing so, we introduce a general S-matrix, with controlled Riemann sheet poles, to generate the teaching dataset. 
Our present method achieves two significant improvements - the inclusion of energy resolution in the DNN design and the utilization of amplitude's error bars in the final DNN inference. 
The simulated energy resolution, which is done by using random energy points, introduces noise in the teaching dataset.
We find that the presence of noise requires the curriculum method to initiate the learning process. 
As a concrete example, we treat the GW-SAID elastic $\pi N$ scattering analysis in  Ref.~\cite{SAIDpiN,SAIDpiNold,GW_SAID} as the experimental data. 
It must be emphasized that our method can only determine the Riemann sheets where the poles are located, but the method does not tell their positions.  
Our approach can be extended to general resonance analysis, not only in hadron physics but also in nuclear or atomic physics, where low-energy scattering reveals interesting universal features.

%\newpage
\mytitle{S-matrix poles}The exact form of S-matrix is not accessible due to the non-perturbative nature of strong interaction. Nevertheless, there are several general properties that we can use. First, causality demands that the scattering amplitudes are analytic in the physical energy sheet~\cite{Kampen1953,EdenAnalytic}. More precisely, the amplitude should have no poles on the physical sheet except below the lowest threshold on the real energy axis where we can have a bound state. Second is the unitarity, which is a consequence of probability conservation. Unitarity helps us to restrict the form of the S-matrix. The last is hermiticity, where the S-matrix is expected to be real below the lowest threshold. The well-known consequence of hermiticity is the reflection principle~\cite{Taylor}, i.e., poles come in conjugate pairs. In the following discussion, we use analyticity, unitarity, and hermiticity in constructing the S-matrix.

Let us consider the two-hadron scattering with two channels where the relative momentum of the $i$th channel ($i=1,2$) is $p_i$, with reduced mass $\mu_i$, and threshold at $T_i$. Specifically, we set the $\pi N$ system as channel 1 and the $\eta N$ as channel 2. 
Aside from the $\pi N$ and $\eta N$ thresholds, there are also the $K\Lambda$ and $K\Sigma$ as shown in Fig.~\ref{fig:SAID_piN}. We justify the two-channel analysis as follows. On the one hand, there is no significant threshold effect with the opening of the $K\Lambda$ channel and, thus, can be ignored~\cite{DCCKNLS2013}. On the other hand, we can avoid the possible effect of $K\Sigma$ channel~\cite{Kaisser1995} by not going beyond the $K\Sigma$ threshold. Hence, it suffices to consider two channels for the present study.

Now, the non-relativistic relation between the scattering energy $E$ and channel momentum $p_i$ is
\begin{equation}
	E = \dfrac{p_i^2}{2\mu_i}+T_i.
	\label{eq:nonrelE}
\end{equation}
The non-relativistic treatment will suffice in our present purpose since we restrict our analysis around the second threshold. The smooth variation of the lower channel can be approximated by Eq.~\eqref{eq:nonrelE} where $p_1$ serves as a label to $E$ for the lower channel. 
Using analyticity, unitarity, and hermiticity in the relevant energy region~\cite{Newton1961,LeCouter1960}, the lower channel S-matrix takes the form
\begin{equation}
	S_{11}(p_1,p_2)=\dfrac{D(-p_1,p_2)}{D(p_1,p_2)}
	\label{eq:smatrix}
\end{equation}
with the scattering amplitude $T_{11}(p_1,p_2)$ obtained via $S_{11}=1+2iT_{11}$.
The function $D(p_1,p_2)$ contains all the pole singularities of $S_{11}(p_1,p_2)$.

We shortly describe the topology of the different Riemann sheets. For the single-channel case, we have the physical (top $[t]$) energy sheet where the imaginary part of the momentum is positive and the unphysical (bottom $[b]$) sheet for the negative imaginary part of the momentum. Here, the upper half of the top sheet is connected to the lower half of the bottom sheet above the threshold. For the two-channel case, we can use the intuitive notation in Ref.~\cite{PearceGibson} where the Riemann sheet is labeled as $[s_1 s_2]$ where $s_i=\{t,b\}$ is the sheet of the $i$th channel. The relevant region of each Riemann sheet is shown in Fig.~\ref{fig:architecture}(c). The interface between the upper half of $[tt]$ and the lower half of $[bt]$ is the real energy axis between the lowest threshold $T_1$ and the higher $T_2$. On the other hand, the lower half of the $[bb]$ sheet is connected to the upper half of $[tt]$ along the real axis above the $T_2$ (dashed lines in Fig.~\ref{fig:architecture}(c)). The $[tb]$ sheet is not directly connected to the physical $[tt]$ sheet but can still leave a noticeable enhancement at the $T_2$. All the Riemann sheets are disconnected below the lowest threshold.

In the generation of simulated amplitudes, we have to consider the general case of more than one pole. 
To do this, we express $D(p_1,p_2)$ as a product of independent $D_j(p_1,p_2)$ such that $D_j(p_1,p_2)=0$ is satisfied by one assigned $j$th pole $E=E^{(j)}_{\text{pole}}$ on a chosen Riemann sheet. Here, the assigned pole $E^{(j)}_{\text{pole}}$ can be one of the poles related to the observed peaks in the scattering region.
The form of $D_j(p_1,p_2)$ can be deduced by hermiticity, where the conjugate pair of energy poles is implied. 
That is, if $E=E^{(j)}_{\text{pole}}$ is a solution to $D_j(p_1,p_2)=0$ on one Riemann sheet, so is its complex conjugate $E=E^{(j)*}_{\text{pole}}$ on the same sheet. 
Thus, we can write $D_j(p_1,p_2)$ in a form where the conjugate partner is explicitly included, and the Riemann sheet is transparently shown:
\begin{equation}
	\begin{split}
		D_j(p_1,p_2)=&\left[(p_1-i\beta_1^{(j)})^2-\alpha_1^{(j)2}\right] \\
		&+\lambda\left[(p_2-i\beta_2^{(j)})^2-\alpha_2^{(j)2}\right]
		\label{eq:Djp1p2}
	\end{split}
\end{equation}
where the absolute values of $\alpha_i^{(j)}$ and $\beta_i^{(j)}$ for $i=1,2$ are determined by the assigned $j$th energy pole $E=E^{(j)}_{\text{pole}}$ through Eq.~\eqref{eq:nonrelE}. The Riemann sheet of $E^{(j)}_{\text{pole}}$ can be chosen arbitrarily by choosing the signs of $\beta_1^{(j)}$ and $\beta_2^{(j)}$. The parameter $\lambda$ is chosen to control the other set of conjugate solutions to $D_j(p_1,p_2)=0$ (called the shadow pole) so that they will not appear on the physical sheet and will remain far from the relevant scattering region. This ensures that $E^{(j)}_{\text{pole}}$ is the only nearby pole produced by $D_j(p_1,p_2)$. For a more detailed discussion of shadow pole related to Eq.~\eqref{eq:Djp1p2}, we refer the  reader to the companion paper in Ref.~\cite{SombilloPRC2021}.

\begin{figure*}[ht!]
	\includegraphics[width=0.95\linewidth]{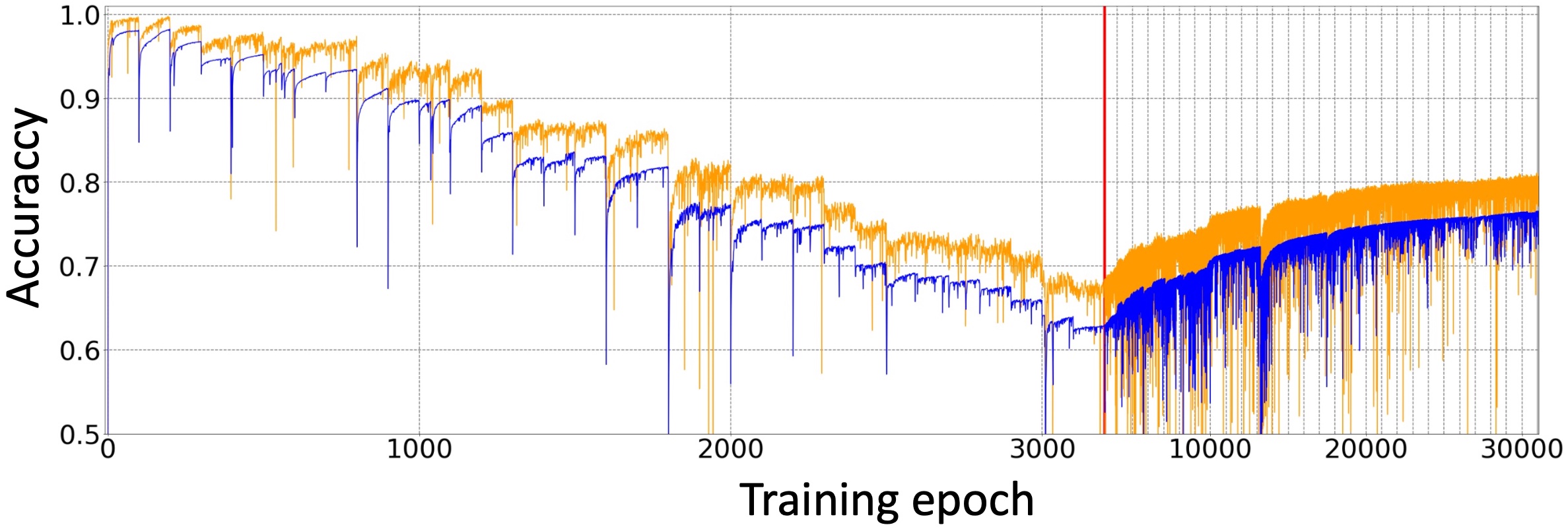}
	\caption{Training (blue) and testing (orange) performance of our DNN model. Curriculum learning is used until epoch 3200 (red vertical solid line). The scale of horizontal axis is changed to epoch/20 after the curriculum learning at the vertical red line.}
	\label{fig:pretraining}
\end{figure*}

\mytitle{Development of DNN}The parametrization that we have introduced allows us to generate simulated amplitudes on arbitrary energy points. However, it is always the case that the experimental data has energy resolution and error bars for the amplitude. The resolution can be accommodated in the design of DNN as follows. We start by determining the region of interest, which in this study is from $\pi N$ to the $K\Sigma$ thresholds. The GW-SAID data has 37 points within the mentioned region, so we split the $\pi N$-$K\Sigma$ into 37 bins. Anticipating that the energy points may not be equally spaced, we randomly select a representative energy point in each bin using a uniform probability distribution. Then, we calculate the amplitude on the randomly picked energy points. Proper labels can be assigned to each simulated amplitude because we have complete control of the poles. The amplitude's error bars are utilized in the last stage of deep learning analysis.

The relevant structures in the GW-SAID data are seen only on some finite interval of the scattering region, as shown in Fig.~\ref{fig:SAID_piN}.
It is, therefore, practical to count only the poles that are within the range $T_2-50\leq \text{Re} E_{\text{pole}}\leq T_2+200$, below the real axis $(-200\leq \text{Im} E_{\text{pole}}< 0$) of the $[bt]$ or $[bb]$ and poles above real axis ($0<\text{Im}E_{\text{pole}}\leq200$) of the $[tb]$. 
Here, the energy units are in MeV.
We do not have to count the conjugate poles because they do not correspond to different independent states. Also, remote background poles are randomly added, by inserting extra $D_{\ell\neq j}(p_1,p_2)$ factor in $D(p_1,p_2)$, but are also not counted. In addition, we restrict the maximum number of nearby poles in any unphysical sheet to four since there are only two prominent structures in the $\pi N$ scattering amplitude. The constraint gives us a total of 35 possible configurations ranging from no nearby poles to at least one pole in each unphysical Riemann sheet. 
We randomly generate poles inside the counting region to produce an equal number of simulated amplitudes per configuration, giving us a total of around $1.8\times 10^6$ simulated amplitudes for the teaching dataset. An independent $3.5\times 10^4$ set of simulated amplitudes is produced to measure the performance of the DNN and to check for possible overfitting.

Figure~\ref{fig:architecture} shows the schematic diagram of our DNN model's operation. Given an input amplitude in Fig.\ref{fig:architecture}(a), our DNN will take the random energy points and the amplitude's real and imaginary parts as the $37\times3$ input-node values. The output nodes are configured to match the total number of possible pole configurations. If the weights and biases (lines in Fig.~\ref{fig:architecture}(b)) are already optimized, then the DNN can give its output, which is schematically represented in Fig.~\ref{fig:architecture}(c). It must be emphasized that our DNN can only count the number of poles in each Riemann sheet and not give the exact pole positions. Now, the performance of our DNN model can be improved in a training loop. Here, all the simulated amplitudes are fed in a forward pass to estimate the cost function, and some variant of stochastic gradient descent is used to do the backpropagation. We trained six different architectures and found that none of them learn the classification problem. In particular, after around 500 epochs, the training accuracies are only around $2.86\%$, which is the accuracy if one makes a random guess out of the 35 possibilities. Our preliminary numerical experiment suggests that our classification problem is too complex for our DNN models, perhaps due to the noise present in the dataset. Thus, to initiate the learning process, we use the curriculum method.

\begin{table}[h!]
	\centering
	\caption{
		Chosen DNN Architecture}
	\begin{tabular}{lcc}
		\hline
		\textbf{Layer} 						 &
		\textbf{Number of nodes}	&        	
		\textbf{Activation Function}  \\
		\hline
		Input 										&
		111+1   										&
		\\
		1st   										&
		200+1   									  &
		ReLU      								 \\
		2nd   									&
		200+1   									 &
		ReLU      							   \\
		3rd   									&
		200+1   									 &
		ReLU      							   \\
		Output   								&
		35   										&
		Softmax      							\\
		\hline							    
	\end{tabular}
	\label{tab:archi}
\end{table}
The basic idea of the curriculum method is to recognize the easy examples of a complex dataset and use them for initial training of the model~\cite{ELMAN199371}. For most classification problems, an additional procedure is expected since one has to identify which part of the dataset is less complicated than the other \cite{CurriculumLearning0,CurriculumLearning1,CurriculumLearning2}. For our case, an additional procedure is needed. We choose the at-most-1-pole configuration as the easy dataset and slowly add multi-pole classifications until all the datasets are introduced. We measure our models' performance on the easy classification set and then devote our computing resources to the best-performing model for the complete curriculum learning~\cite{SombilloPRC2021}. We have chosen the DNN model with specifications given in Table~\ref{tab:archi}. All numerical calculations related to the design of DNN and training are done using the define-by-run framework of Chainer \cite{Chainer2015,Chainer2017,Chainer2019,ChainerGithub}.

Fig.~\ref{fig:pretraining}. shows the performance of our chosen model during and after the curriculum learning. Except for the easy dataset, one new classification is added every after 100 epochs. We also vary the mini-batch size in each epoch as we saw fit. The smaller mini-batch size is optimal in the early part of the new-classification-epoch while the large ones are better for the later part. The sudden accuracy drop corresponds to the introduction of a new classification. After 3,200 epochs, we now have a pre-trained DNN model that can detect up to four nearby poles on any Riemann sheet with training accuracy of $63.5\%$ and testing accuracy of $68.3\%$. We further train the model up to 31,050 epochs, obtaining a final performance of $76.5\%$ for the training and $80.4\%$ for the testing.

\mytitle{Results and Discussion}The GW-SAID amplitude in Fig.~\ref{fig:SAID_piN} is used as the experimental data in the following discussions.
Note that the proper description of experimental data must include the uncertainties, and the deep learning approach is no exception here. 
In addition, we should be able to estimate the confidence of the trained DNN's prediction once the experimental data is fed. 
To accomplish these, we first interpret the error bars of the amplitude as a collection of points weighted by some probability distribution. 
The most natural choice is to treat each error bar as one standard deviation of a Gaussian distribution. 
We produce a collection of $10^6$ amplitudes by combining the points in each error bar. Finally, the amplitudes are fed directly to the trained DNN models, and the outputs are counted. The output with the highest count is the possible pole configuration of the experimental data.

\begin{table}[ht!]
	\centering
	\caption{Result of the DNN inferences on the GW-SAID $\pi N$ scattering amplitude. The error bar is interpreted as one standard deviation to generate $10^6$ amplitudes from the experimental data.}
	\begin{tabular}{c|c|c|c}
		\hline
		\textbf{Percentage}&
		\textbf{bt}&
		\textbf{bb}&
		\textbf{tb} \\
		\hline
		44.6\%&
		1&
		1&
		2\\
		34.1\%&
		1&
		1&
		1\\		
		16.4\%&
		0&
		1&
		3\\	
		4.9\%&	
		0&
		1&
		2\\
		\hline		
	\end{tabular} 
	\label{tab:results}
\end{table}	

Out of the 35 possibilities, our trained DNN linked only four-pole configurations to the $\pi N$ scattering amplitude. 
As shown in Table~\ref{tab:results}, the structures in Fig.~\ref{fig:SAID_piN} are due to one pole in $[bt]$, one pole in $[bb]$ and at most two poles in $[tb]$. 
Note that no assumptions are made on any of the detected poles since they are produced independently in the training dataset.
It is now up to some dynamical model to interpret their origin and explain how they relate to each other.
Nevertheless, we can make some general remarks.

The $[bb]$ pole is consistently identified by the DNN in all generated amplitudes. 
Note that the peak well above the $\eta N$ threshold can only be caused by a pole in the $[bb]$ sheet. Thus, the detected $[bb]$ pole can be unambiguously associated with the enhancement between the $K\Lambda$ and $K\Sigma $ thresholds. 

It seems likely that the detected $[bt]$ pole is linked to the enhancement around the $\eta N$ threshold, but this can only happen if the peak is close to the unitarity limit, i.e., $\text{Im}T_{11}\sim 1$. The fact that $\text{Im}T_{11}<1$ around the $\eta N$ threshold suggests that the $[bt]$ pole is far but still within the counting region. Furthermore, comparing the configurations with $44.6\%$ and $16.4\%$ percentages in Table~\ref{tab:results}, implies that 
the $[bt]$ pole is sometimes interpreted as a $[tb]$ pole by our DNN. This is possible if the $[bt]$ pole is close to the $[bt]$-$[tb]$ interface which is above the $\eta N$ threshold (branch cut above $T_2$ in Fig.~\ref{fig:architecture}c). Now, the origin of the $[bb]$ and $[bt]$ poles discussed so far can only be deduced using a dynamical model~\cite{PearceGibson,Badalyan1982}. Nevertheless, we can speculate that the detected $[bb]$ and $[bt]$ poles, which are both above the $\eta N$ threshold, might be associated with the enhancement observed between the $K\Lambda$ and $K\Sigma $ thresholds.

There will always be a cusp at the S-wave $\eta N$ threshold regardless of whether there is a pole. However, the cusp can only be enhanced significantly in the presence of at least one pole. 
With $[bb]$ and $[bt]$ poles already allocated to the enhancement between $K\Lambda$ and $K\Sigma $ thresholds,
we are only left with at most two poles in the $[tb]$ sheet to associate with the $\eta N$ peak. 
The peak of the $\mbox{Im} T_{11}$ above the $\eta N$ is an indication that one of the detected $[tb]$ poles is close to the $[tb]$-$[bb]$ interface below the $\eta N$ threshold.
It is possible that one of the $[tb]$ poles is originally from the $[bb]$ sheet and was drawn to its present position due to strong $\eta N$ channel coupling. 
Again, this last statement must be further verified by a dynamical model.

Our approach can go beyond the conventional model-fitting scheme, where the error bar is typically interpreted as one standard deviation of a Gaussian distribution. In the companion paper Ref.~\cite{SombilloPRC2021}, we show that the choice of a probability distribution for the description of error bars does not affect the outcome of DNN inference. In particular, the same set of configurations is obtained, and the configuration with the highest count remains no matter which error bar interpretation is used. This observation demonstrates that our deep learning approach gives a statistically robust interpretation of experimental data.

\mytitle{Conclusion and Outlook}In this letter, we have shown for the first time how to implement deep learning in the study of coupled-channel scattering with near-threshold enhancement. The DNN model is designed to accommodate the energy resolution of the experimental data and the error bars, which are utilized in the final DNN inference. As a result, the confidence of DNN's inference can be estimated by counting the output, with the highest count being the best description of experimental data.

It must be pointed out that the inclusion of uncertainty comes with a price. That is, we have to inevitably deal with the presence of noise in the training dataset. Nevertheless, we have shown that a simple curriculum method can initiate the learning process. Here, an easy classification set is introduced first to the DNN, and then the more complicated ones are added as the training progresses. It is, therefore, imperative to adopt a curriculum-type of training if we wish to include the experimental uncertainties.

The deep learning approach used in this letter can be extended to any peak analysis of the experimental data. 
In the present study, we focused only on identifying the pole configuration that best describes a near-threshold enhancement.
No a priori assumption is made on the detected poles; they are independently produced in the generation of the teaching dataset.
However, to further identify the nature of near-threshold phenomena, we need to employ a suitable dynamical model to reproduce the detected poles. 
We can then determine the origin of enhancements by using the obtained pole configuration and tracing their trajectories through the coupling mechanism of a dynamical model.

\mytitle{Acknowledgment}
This study was supported in part by MEXT as “Program for Promoting Researches on the Supercomputer Fugaku” (Simulation for basic science: from fundamental laws of particles to creation of nuclei).
DLBS is supported in part by the DOST-SEI ASTHRDP postdoctoral research fellowship. 
YI is partly supported by JSPS KAKENHI Nos. JP17K14287 (B) and 21K03555 (C).
AH is supported in part by JSPS KAKENHI No. JP17K05441 (C) and Grants-in-Aid for Scientific Research on Innovative Areas, No. 18H05407 and 19H05104.

\bibliography{mybib}

\end{document}